\documentclass[pre,amsmath, twocolumn, showpacs, superscriptaddress]{revtex4-1}
\usepackage{graphicx}

\begin{document}

\title{Nonequilibrium fluctuation theorem for systems under discrete and continuous feedback control}

\author{Anupam Kundu} 
%\affiliation{ ESPCI, Paris, France, 75231}
\affiliation{ PCT-UMR, CNRS, Gulliver 7083, ESPCI, 10 rue Vauquelin, F-75231 Paris, France}

\begin{abstract}
Without violating causality, we allow performing measurements in time reverse process of a feedback manipulated stochastic system. As a result we come across an entropy production due 
to the measurement process. This entropy production, in addition to the usual system and medium entropy production, constitutes the total entropy production of the combined system of the reservoir, the system and the feedback controller. 
We show that this total entropy production of "full" system satisfies an integrated fluctuation theorem as well as a detailed fluctuation theorem as expected. We illustrate and verify this idea through 
explicit calculation and direct simulation in two examples.
\end{abstract}

\vspace{0.5cm}
%\date{\today}

%\pacs{: }
\maketitle
\section{Introduction}
In contrast to the linear response results, recently a few exact predictions for systems far from equilibrium have been discovered.  Under the common guideline these predictions are called \emph{Fluctuation theorems}(FT) 
\cite{eva93,eva94,jar97,cro98, cro99}. 
In 1997 C. Jarzynski proposed an exact equality between the work ($W$) performed on system and the free energy difference ($\Delta F$) \cite{jar97}:
\begin{equation}
\langle e^{-\beta (W-\Delta F)} \rangle = 1 
\label{Jarz}
\end{equation} 
from which one gets the expected inequality of the second law of thermodynamics and the fluctuation-dissipation theorem. A further generalization and detailed study along the same line was given by Crooks and others. These studies are 
  about the time-reversal symmetry of the microscopic dynamics and commonly termed as the \emph{detailed fluctuation theorem} .
The detailed fluctuation theorem relates the probability to observe a microscopic trajectory of the system undergoing forward process to the probability to observe the time reversed trajectory in the 
reverse process. The concept of relating forward and reverse process trajectory probabilities has been beneficial for defining the entropy productions in the system and understanding the irreversibility at microscopic scale. These relations are very important in understanding thermodynamics of small systems and widely used for calculating equilibrium free energy differences from nonequilbrium processes, both in experiments as well as in computer simulations.

More recently, a new interest has been developed in studying FTs for \emph{feedback} manipulated systems in which, one performs a measurement to get some information about the system and then uses that information to control the dynamics further. In biological molecular machines and nano machines one naturally uses feedback control. Also the study of feedback control of stochastic system plays an important role in understanding the second law of thermodynamics. In this spirit there have been many studies in the "Maxwell demon" type set up \cite{sag08,sag10,sag12,bar00}. Naturally, usual FT relations has to be extended for systems under feedback control. The first work 
along this line was reported by T. Sagawa and M. Udea \cite{sag10,sag12} in which, they have shown that the Jarzynski relation for feedback controlled systems would be extended to 
\begin{equation}
\langle e^{-(\beta W +I-\beta \Delta F)} \rangle = 1
\end{equation} 
where $I$ is the amount of information obtained through the measurements performed in the forward process. Same relation for entropy production was discussed by D. Abreu and U. Seifert in \cite{sei12}. Recently, Crooks type studies have also been attempted in these kind of systems \cite{hor10,pon10,jayan12}. In all these studies the possibility of entropy production due to the interaction between the system and the measurement apparatus was not well considered except 
the study by Cao and Feito \cite{cao09} in which, they have given a way to compute the entropy reduction in feedback controlled system due to repeated operation of controller. It is quite natural to ask, what would be the total amount of entropy production  $\Sigma$ of the "full" system (the system + reservoir + feedback controller) ? Can this entropy be included in the framework of Fluctuation theorems? It is clear that, along with the usual medium entropy production and the system entropy production one should have a contribution from the measurement process in $\Sigma$. To identify an entropy production one generally compares the path probabilities of forward and reverse process. In \cite{hor10}, J. Horowitz and S. Vaikunthanathan have shown a way to construct reverse process of a given forward process in a feedback manipulated system. From their construction of reverse process, it is not possible to identify such a quantity, which has the properties of entropy production, because, they have excluded the possibility of performing measurements in the reverse process in order to respect causality. In this paper, we show that, at least for a specific class of forward processes, a corresponding reverse process can be constructed in which one can allow performing measurements without violating causality. This construction of the reverse process allows one to get an entropy production due to the measurement process, along with the other two entropy productions (medium and system entropy productions). With this identification in the ratio of trajectory probabilities of the forward and reverse process, we show that the extra contribution in entropy production due to control process can be included to verify a \emph{detailed fluctuation theorem} for the closed "full" system.

In sec.~(\ref{deriv}) we describe the way to construct the reverse process and using this we derive the different Fluctuation theorems. Next in sec.~(\ref{Illus}) two examples are studied in detail to 
demonstrate the Fluctuation theorems obtained in sec.~(\ref{deriv}). First, in subsection (\ref{Lange-particle}), we describe pulling of an overdamped Langevin particle manipulated through discrete feedback control and then we discuss
dynamics of Brownian particle under continuous feedback control in subsection (\ref{kimmodel}). We end by giving our conclusions in sec.~(\ref{conclu}).

%%%%%%%%%%%%%%%%%%%%%%%%%%%%%%%%%%%%%
\section
{Derivation} \label{deriv}
\begin{figure}[t]
\includegraphics[scale=.35]{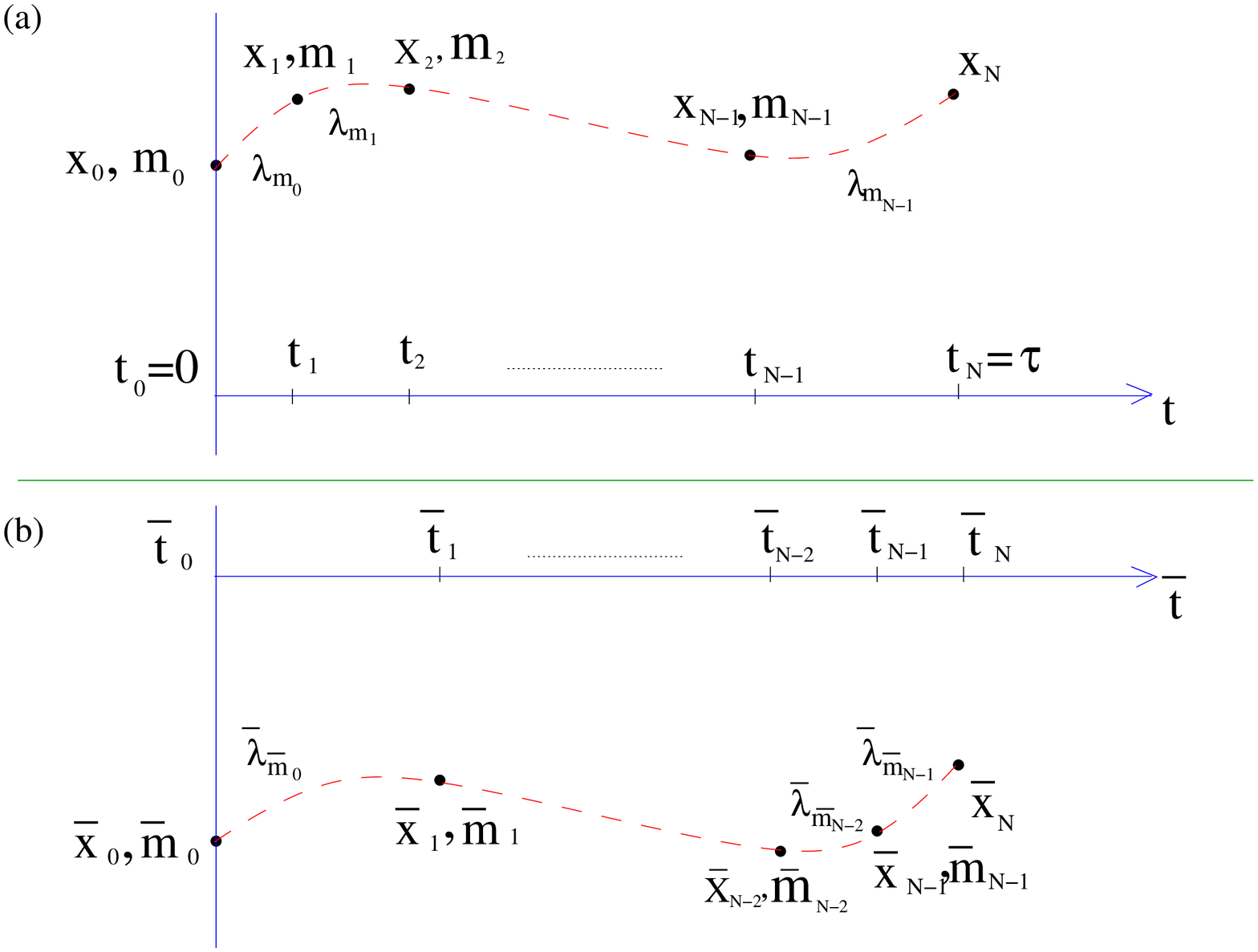}
\caption{(Color online) Illustration of the forward trajectory (a) and reverse trajectory (b) with measurements and feedback loops. The pair $(x,m)$ next to each black dot on trajectories (red dashed curves)
 corresponds to state of the system and measurement outcome, respectively. $\lambda_m$ between two black dots represents the protocol between times corresponding to that two dots.}
\label{fig1}
\end{figure}

Let us consider driving a stochastic system through closed loop feedback control from $t=0$ to $t=\tau$. 
At each predetermined time $t_i;~i=0,~1,~2,...(N-1)$ a measurement is performed on the system. We consider each measurement to be imprecise and model them by introducing a conditional probability $p(m|x)$ of getting outcome $m$ while the actual state is $x$. First we choose the system in state $x_0$ with probability $P^F_{in}(x_0)$ at $t=0$. Then we make 
a measurement on the system and get an outcome $m_0$ with probability $p(m_0|x_0)$. We start manipulating the system with a protocol $\lambda_{m_0}(t)$ (dependent on $m_0$) and manipulate it till time $t_1$ where, we make the second measurement. The system reaches the state $x_1$ at time $t_1$ with transition probability $W(x_1,t_1|x_0,t_0;\lambda_{m_0}(t))$ and an outcome $m_1$ is obtained after second measurement, with probability $p(m_1|x_1)$. 
We now modify our protocol from 
$\lambda_{m_0}(t)$ to $\lambda_{m_1}(t)$ and run it from $t_1$ to $t_2$.  We continue performing measurements and modify protocols at the predetermined times $t_i;~i=0~\text{to}~N-1$ which 
can be seen in fig.~(\ref{fig1}). Here we assume that for each measurement outcome $m $ there is a unique protocol $\lambda_{m}(t)$ and it is completely independent of the outcomes obtained in the earlier measurements \emph{i.e.} 
$\lambda_{m_1}(t)$ is independent of $m_0$, $\lambda_{m_2}(t)$ is independent of $\{m_0,m_1\}$ and so on.
% \emph{i.e.} the protocols at successive time intervals are uncorrelated. 
In general, the value of the overall protocol has a discontinuous jumps at times $t_i$ and the amount of the jump is $|\lambda_{m_{i-1}}(t_{i-1})-\lambda_{m_i}(t_i)|$. The above procedure is called the \emph{forward process}. In this construction of
forward process, dynamics of a Markovian system remains always Markovian even though it is manipulated through feedback. Let the collection of the states $x_i=x(t_i)$'s and 
the outcomes $m_i$'s of the measurements at times $t_i$, are represented by $\mathcal{X}=\{x_i;~\forall~ i=0...N\}$ and 
$\mathcal{M}=\{m_i;~\forall~ i=0...(N-1)\}$, respectively. The joint probability 
$\mathcal{P}_F(\mathcal{X},\mathcal{M})$ of observing a path $(\mathcal{X},~\mathcal{M})$ in the forward process is explicitly given by 
\begin{eqnarray}
\mathcal{P}_F(\mathcal{X},\mathcal{M})&=& \prod_{i=0}^{N-1}W(x_{i+1},t_{i+1}|x_i,t_i;~\lambda_{m_i}(t)) 
\nonumber \\ && ~~~~~~~\times p(m_i|x_i)P_{in}^F(x_0). \label{path-pro-F}
\end{eqnarray}

A \emph{reverse process} is constructed the same way as the above forward process but the measurements are performed at reverse times $\bar{t}_i=\tau - t_{N-i};~i=0~\text{to}~N-1$. Time in the reverse process is denoted by $\bar{t}$ and it is related to the time in forward process as : $\bar{t}=\tau -t$. Also the state and outcome variables in the reverse process are represented with "$(\bar{..})$"s. At time $\bar{t}_0$, let us start from a 
state $\bar{x}_0$ of the system, which is chosen with probability $P_{in}^R(\bar{x}_0)$.   We make a measurement and get an outcome $\bar{m}_0$ with conditional probability $p(\bar{m}_0|\bar{x}_0)$. Here we assume that the measurement process is 
identical in both the forward and reverse processes.  Now corresponding to 
this particular outcome $\bar{m}_0$ there must be a unique protocol $\lambda_{\bar{m}_0}(t)$ in the forward process. We take that protocol and run it in reverse from $\bar{t}_0$ to $\bar{t}_1$.  
The resultant protocol in reverse process is represented by $\bar{\lambda}_{\bar{m}_0}(\bar{t}) = \lambda_{\bar{m}_0}(\tau-t)$. Then at time $\bar{t}_1$ we perform a second measurement and if the outcome is $\bar{m}_1$ we run the protocol 
$\bar{\lambda}_{\bar{m}_1}(\bar{t}) =  \lambda_{\bar{m_1}}(\tau-t)$ from $\bar{t}_1$ to $\bar{t}_2$. As in the forward process, we continue performing measurements at times $\bar{t}_i;~i=0$ to $N-1$ and modify the protocols accordingly. 
%After that we evolve the system with protocol $\bar{\lambda}_{\bar{m}_{(N-1)}}(t)$ till time $\bar{t}_N$. 
The joint probability of observing a trajectory 
$\bar{\mathcal{X}}=(\bar{x}_0,~\bar{x}_1,....,\bar{x}_N)$  and an outcome trajectory $\bar{\mathcal{M}}=(\bar{m}_0,~\bar{m}_1,.....,\bar{m}_{(N-1)})$ in reverse process is given by 
\begin{eqnarray}
\mathcal{P}_R(\bar{\mathcal{X}},~\bar{\mathcal{M}}) &=& \prod_{i=0}^{N-1}W(\bar{x}_{i+1},\bar{t}_{i+1}|\bar{x}_i,\bar{t}_i;~\bar{\lambda}_{\bar{m}_i}(t)) \nonumber \\ 
&& ~~~~~~~\times p(\bar{m}_i|\bar{x}_i){P}_{in}^R(\bar{x}_{\bar{t}_0}), \label{path-pro-R-n}
\end{eqnarray}
where, $\bar{x}_i=\bar{x}(\bar{t}_i)$. We can easily see that, for each path $(\mathcal{X},~\mathcal{M})$ in the forward process,  
one can have a conjugate path $(\bar{\mathcal{X}},~\bar{\mathcal{M}})$ in the reverse process such that 
%\begin{equation}
$\bar{x}_i=x^*_{N-i}~~\text{and}~~ \bar{m}_i=m_{N-i-1}.$
%\end{equation} 
Here $*$ represents sign change of odd quantities under time reversal.
%Following earlier works in the literature, we naturally define the ratio $\mathcal{R}$ of the  path probabilities of a forward path and it's conjugate path as :
%\begin{equation}
%\mathcal{R}=\frac{\mathcal{P}_F(\mathcal{X},\mathcal{M})}{\mathcal{P}_R(\bar{\mathcal{X}},\bar{\mathcal{M}})} .
%\end{equation} 

When we take the initial distribution in the reverse process to be same as the final distribution of the forward process \cite{cro98, sei05,harr07, chern06} \emph{i.e.}  $\bar{P}_{in}^R(x) = P_{fn}^F(x)$, then from the ratio of 
the path probabilities of a forward path and it's conjugate path we get 
\begin{equation}
\frac{\mathcal{P}_F(\mathcal{X},\mathcal{M})}{\mathcal{P}_R(\bar{\mathcal{X}},\bar{\mathcal{M}})} =e^{(\Delta s_s + \Delta s_m )} \times~\prod_{i=0}^{N-1}\frac{p(m_i|x_i)}{p(\bar{m}_i|\bar{x}_i)}, 
\label{DFT4agn}
\end{equation}
where, the medium entropy production $\Delta s_m$ and the system entropy production $\Delta s_s$ are, respectively, given by \cite{espo10} 
\begin{eqnarray}
&&\Delta s_m = ln \Big[\prod_i^{N-1} \frac{W(x_{i+1},t_{i+1}|x_i,t_i;~\lambda_{m_i}(t))}{W(x^*_i,\tau-t_i|x^*_{i+1},\tau-t_{i+1};~\lambda_{m_i}(\tau-t))}\Big], \nonumber\\
&&\Delta s_s = ln \Big[ \frac{P_{in}(x_0)}{P_{fn}(x^*_N)} \Big].
\label{entro-def}
\end{eqnarray}
We identify the rest part on the \emph{r.h.s} of eq.~(\ref{DFT4agn}) as another entropy production $\Delta s_p$  and write  
\begin{eqnarray}
&&\Delta s_p= ln \Big[\frac{p(m_0|x_0)~p(m_1|x_1)....p(m_{N-1}|x_{N-1})}{p(m_{N-1}|x^*_{N})~p(m_{N-2}|x^*_{N-1})....p(m_0|x_1^*)} \Big ]. \label{feed-entro}
\end{eqnarray}
Here we recall that, the measurement process was idealized by introducing a conditional probability $p(m|x)$ of getting an outcome $m$ given that the system was in state $x$. 
At this point, one can think of the outcome $m$ as some "representation" of the state of the measurement apparatus and $\Delta S_p$ as the measure of the disorder, created in the full system through {\it{imprecise}} measurements. 
Under certain assumptions, this entropy production can be thought of as difference between the amounts of information acquired in the forward process and the reverse process. 
Considering this entropy into account one can write an expression for the total entropy of the "full" system as 
\begin{equation}
\Sigma = \Delta s_s + \Delta s_m + \Delta s_p,
\end{equation}
and hence the eq.~(\ref{DFT4agn}) is re-written as 
\begin{eqnarray}
&&\frac{\mathcal{P}_F(\mathcal{X},\mathcal{M})}{\mathcal{P}_R(\bar{\mathcal{X}},\bar{\mathcal{M}})} =e^{\Sigma}.
\label{DFT5agn}
\end{eqnarray}
From the above relation it is easy to see the integrated fluctuation theorem 
$\langle e^{-\Sigma} \rangle =1$ and consequently the second law of thermodynamics for the "full" system
\begin{equation}
\langle \Sigma \rangle \geq 0. \label{secondlaw}
\end{equation}
In the following scenario, we consider that, in each realization of the forward process, the system was being manipulated by a constant protocol of value $\lambda_{in}$ till time $t=0$ starting from $t=-\infty$ 
and after time $\tau$ the system is kept under another constant protocol $\lambda_f$ till time $t=\infty$. For the reverse process we imagine the opposite. We can think that the initial distribution of the forward (reverse) process 
to be a steady state distribution corresponding to $\lambda_{in}$($\lambda_f$). In this situation, we can interpret $\Sigma_F[\mathcal{X},~\mathcal{M}]$ and 
$\Sigma_R[\bar{\mathcal{X}},~\bar{\mathcal{M}}]$ as entropy productions in the forward and reverse process, respectively, and  from eq.~(\ref{DFT5agn}) we get the following detailed fluctuation theorem \cite{harr07, chern06} 
\begin{equation}
\frac{Prob(\Sigma_F[\mathcal{X},~\mathcal{M}]=\Sigma)}{Prob(\Sigma_R[\bar{\mathcal{X}},~\bar{\mathcal{M}}]=-\Sigma)}=e^{\Sigma}. \label{DFT5}
\end{equation} 
This is a detailed fluctuation relation for a feedback controlled system. The corresponding characteristic functions of the forward and reverse probabilities, denoted respectively by $Z_F(u)$ and $Z_R(u)$, satisfy the symmetry 
\begin{equation}
Z_F(u)=Z_R(1-u) \label{gDFT}
\end{equation}
 where 
$Z_{F,R}(u)=\int_{-\infty}^{\infty} e^{-u \Sigma}P_{F,R}(\Sigma) d\Sigma$. 
Under appropriate assumptions, it is easy to extend the above analysis for a quantum system, coupled to a bath and a feedback controller.  
Recently,  M. Esposito and G. Schaller \cite{espo12} have also used similar feedback mechanism in Kramer's barrier problems where, they change the energy barriers heights between the system states without changing
the system's energy through specific control. They also have obtained same results in their study.

 When the initial and final distributions are equilibrium distributions corresponding to 
$\lambda_{in}$ and $\lambda_{fn}$ and inverse temperature $\beta$ then, we write $\Sigma= \beta (W - \Delta F) + \Delta s_p$ where $\Delta F$ is 
the free energy difference between the two equilibrium states. From the second law in eq.~(\ref{secondlaw}) we get 
\begin{eqnarray}
W &\geq& \Delta F  + \frac{\langle \Delta s_p \rangle}{\beta}  \\
&\geq& \Delta F + \frac{1}{\beta} \sum_{\mathcal{M}}\int \mathcal{D X}~\mathcal{P}_F(\mathcal{X},\mathcal{M})\sum_{i=0}^{N-1}ln \left ( \frac{p(m_i|x_i)}{p(\bar{m}_i|\bar{x}_i)} \right). \nonumber
\end{eqnarray}
%An extra amount of work can be extracted by using the feedback control. 
A similar relation is obtained in \cite{espo11,kawai07} where the authors have shown that an extra amount of work can be extracted 
when one starts and ends with distributions different from equilibrium distributions. This extra work is related to the amount of information required to produce the 
not-equilibrium distributions from known equilibrium distributions.

Now we illustrate and verify the above fluctuation theorems in eq.~(\ref{DFT5}) and (\ref{gDFT}) for two different examples : one with discrete feedback control and the other one with continuous feedback control.

%%%%%%%%%%%%%%%%%%%%%%%%%%%%%%%%%%%%%%%%%%%%

\section{Illustration} \label{Illus}
%{\it{Illustration:}} 

\subsection{Overdamped Langevin particle under discrete feedback control}\label{Lange-particle}
We consider controlled pulling of a harmonically trapped Langevin particle for duration $t=0$ to $\tau$. The corresponding stochastic equation is given by
\begin{equation}
\gamma \dot{x}(t) = -\frac{d U(x,\lambda_y(t))}{dx} + \sqrt{2\gamma T}\eta(t), \label{Lange-eq}
\end{equation}
where $\eta$ is a mean zero, unit variance Gaussian white noise, $x(t)$ is the position of the particle and 
$U(x,\lambda_y(t))=\frac{k}{2}(x - \lambda_y(t))^2$ is the effective potential in which the particle moves. The term $\lambda_y(t)$ represents 
the protocol with which we manipulate the particle. Subscript $y$ shows that the protocol explicitly depends on measurement outcomes. We consider single measurement process. The feedback mechanism is implemented as follows : 
Initially the particle is considered to be in thermal 
equilibrium at temperature $T$. A measurement of the position of the particle is performed at this time. Given that the position of the particle at $t=0$ was $x_0$, we get an outcome $y$ with the following conditional probability : 
$p(y|x)= \frac{1}{\sqrt{2\pi \Delta^2}}e^{-\frac{(y-x)^2}{2 \Delta^2}}$ where $\Delta$ quantifies the error in the measurement. Now we pull the particle by applying a protocol $\lambda_y(t)$ from $t=0$ to $t=\tau$. We consider 
two different protocols %(shown in fig.~(\ref{proto}))
\begin{eqnarray}
(a)~\lambda^a_y(t) &=& \alpha y~\Theta(t)\Theta(\tau-t) \nonumber \\
(b)~\lambda^b_y(t) &=& \alpha y~\sin(\frac{\pi t}{\tau})~\Theta(t)\Theta(\tau-t), \label{protocols}
\end{eqnarray}    
where $\Theta(t)=1$ for $t>0$ and $0$ for $t\leq 0$. A dimensionless parameter $\alpha$ is introduced to control the strength of the protocol. 

%\begin{figure}[t]
%\includegraphics[scale=.25]{protocol.eps}
%\caption{Schematic diagram of the protocols (a) and (b) in eq.~(\ref{protocols}).}
%\label{proto}
%\end{figure}

In the reverse process we once again consider the particle to be in thermal equilibrium at 
temperature $T$. We perform a measurement on the position of the particle at the beginning and suppose that we get an outcome $\bar{y}$ with conditional probability $p(\bar{y}|\bar{x}_0)$. Now we consider the corresponding forward 
protocol $\lambda_{\bar{y}}(t)$ [\emph{e.g.} for case (b) we have $\lambda_{\bar{y}}^b(t)= \alpha \bar{y}~\sin(\frac{\pi t}{\tau})$] and run it in reverse. Note that the protocols chosen for a particular $y$ are time reversal symmetric. 
So the functions 
$P_F(\Sigma)$ and $P_R(\Sigma)$ are same and consequently $Z_F(u)$ and $Z_R(u)$ are same. We denote the probability by $P(\Sigma)$ and the corresponding characteristic function by $Z(u)$. Considering the 
ratio of the joint probabilities of the forward path $(x(t),y)$ and its conjugate reverse path $(\bar{x}(\bar{t}),y)$ we get the total entropy $\Sigma$ 
of the "full" system as follows :
\begin{eqnarray}
(a)~~\Sigma_a &=& \frac{\alpha k y}{T}(x_{\tau}-x_0) + \frac{1}{2 \Delta^2}\left ((y-x_{\tau})^2 - (y-x_0)^2\right )~; \nonumber \\
(b)~~\Sigma_b &=& -\frac{\pi \alpha y}{T\tau}\int_0^{\tau} \sin(\frac{\pi t}{\tau})(x - \alpha y~\sin(\frac{\pi t}{\tau})) \nonumber \\
&&+ \frac{1}{2 \Delta^2}\left ((y-x_{\tau})^2 - (y-x_0)^2\right ). \label{entro-langevin}
\end{eqnarray}
For case (a), we get an analytical expression of $Z_a(u)$ which has the following form : 
% Since the expression of $Z(u)$ in terms of all the parameters ($k,~T,~\Delta,~\alpha,~{\rm{\tau}}$) is too big, we give it's expression for the following parameter values: $k=T=\Delta=1.0$ and $\alpha=2.0$ and it looks like 
\begin{eqnarray}
Z_a(u)&=& \sqrt{\frac{\beta k f_1\Delta^2}{2g_1(u) g_2(u)}},   \label{zofa} 
\end{eqnarray}
where, 
%\begin{eqnarray}
$g_1(u) = 2\sigma_y^2(\Delta^2+u\sigma_{\tau}^2)B(u)$ 
and 
%  \nonumber \\
$g_2(u)=\frac{\beta k f_1}{2}-A(u)-\frac{C(u)^2}{4B(u)}$
%\end{eqnarray}
with $f_1=\frac{T}{T+k\Delta^2}~,~\sigma_y^2=\Delta^2 f_1$ and $\sigma_{\tau}^2=\frac{T}{k}(1-e^{-\frac{2k\tau}{\gamma}})$. The expressions for $A(u)$, $B(u)$ and $C(u)$ are given in Appendix. 
\begin{figure}[t]
\includegraphics[scale=.3]{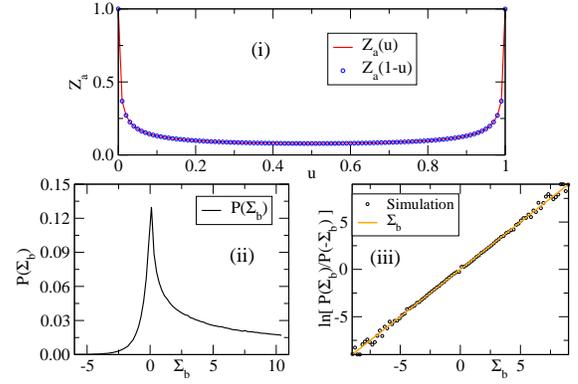}
\caption{(Color online) Plot (i) shows that the characteristic function $Z_a$ of $\Sigma_a$, satisfies symmetry in eq.~(\ref{gDFT}) for protocol $(a)$. Plot (ii) shows the distribution of $\Sigma_b$ during 
forward process of protocol $(b)$ and plot (iii) shows that $\Sigma_b$ satisfies the symmetry in eq.~(\ref{DFT5}). Parameters for this plot are : 
($\tau=2.0,~k=1.2,~\Delta=0.2,~\gamma=1.0,~\beta=0.5$ and $\alpha=5.0$)}
\label{fig2}
\end{figure}
One can easily check in Mathematica that the functions $g_1(u)$ and $g_2(u)$ individually satisfy the symmetry $g(u)=g(1-u)$ for any arbitrary set of parameters and consequently we have $Z_a(u)=Z_a(1-u)$ as shown in subplot (i) of 
fig.~(\ref{fig2}). 

It is difficult to get a closed form expression of $Z_b(u)$. Instead we directly simulate the corresponding Langevin equation using velocity Verlet algorithm and get $P(\Sigma_b)$ in (ii) of fig.~(\ref{fig2}). 
In subplot (iii) of the same figure we plot 
logarithm of the ratio $\frac{P(\Sigma_b)}{P(-\Sigma_b)}$ Vs. $\Sigma_b$ and see that the figure is a straight line with slope $1$. This verifies the relation in eq.~(\ref{DFT5}).

\subsection{Brownian particle under continuous feedback control}\label{kimmodel}
In \cite{kim04,kim07}, Kim and Qian have considered cooling of a molecule using continuous velocity dependent feedback. In their study, they have precise deterministic feedback controller
which can provide drag forces proportional to the instantaneous velocity of the particle. Generally, feedback controller makes a measurement of instantaneous velocity of the particle and provide the drag force according to the outcome 
of the measurement. In practice, every measurement involves errors, which leads to imprecise outcomes and the subsequent control on the dynamics of the system is dependent on these outcomes. This kind of control is called 
nondeterministic feedback control or closed loop feedback control. We here revisit the cooling problem using this control. 
Starting with discrete feedback process, we go to continuous feedback process by taking appropriate limits of the parameters. 

 The dynamics of a 
Brownian particle of mass $m$ in contact with a reservoir of temperature $T$,
 is given by $m\dot{v} = -\gamma v +\eta(t)$, where $v$ is the instantaneous velocity of the particle and $\eta(t)$ is a mean zero white Gaussian noise of strength $2 \gamma T$. 
We assume that the particle was in thermal equilibrium at $t=0$. The particle dynamics is now manipulated through feedback control from $t=0$ to $t=\tau$. 
For convenience, let us divide the total time $\tau$ into $N$ equal intervals of length $\delta t$ and 
label these times by $t_i$ for $i=0~\text{to}~(N-1)$ with $t_0=0$ and $t_N=\tau$. Performing measurements on the velocity of the particle at each time $t_i$($i=0~\text{to}~N-1$), we get an outcome $v'_i$ whose probability is conditional 
on the actual velocity $v_i$ of the particle at time $t_i$ and is given by 
$p(v_i'|v_i)=\frac{1}{\sqrt{2 \pi \Delta ^2}}e^{-\frac{(v_i-v_i')^2}{2\Delta^2}}$.  We apply a constant drag force of amount 
$ -\gamma' v_i'$ in the time interval $t_{i+1}-t_i$. 
Now the equation of motion of the manipulated particle becomes  
\begin{equation}
m\dot{v}= - \gamma v - \gamma' v_i' +\eta(t);~t_i \leq t < t_{i+1}. \label{Blangeq}
\end{equation}
%Restating the above equation as energy conservation equation one gets the work done $\Delta W$ and heat flow $\Delta Q$ as 
%\begin{eqnarray}
%\Delta W &=& - \sum_{i=0}^{N-1} \gamma' \frac{(v_{i+1}+v_i)}{2}v_i' \nonumber \\
%\Delta Q &=& \sum_{i=0}^{N-1} \frac{(v_{i+1}+v_i)}{2} (\gamma v_i \Delta t + \int_{t_i}^{t_{i+1}}\eta(t) dt). 
%\end{eqnarray}
For each different realization of $\eta(t)$, the particle will follow a different trajectory in the phase space along with a different outcome trajectory. For notational convenience, let us represent the particle trajectory 
by the collection of it's velocities at times $t_i~;~i=0..N$ \emph{i.e.} ${\bf{v}}=\{v_0,v_1,...,v_N\}$ and the outcome trajectory by ${\bf{v'}}=\{v_0',v_1',...,v_{N-1}'\}$.
 \begin{figure}[t]
\includegraphics[scale=.35]{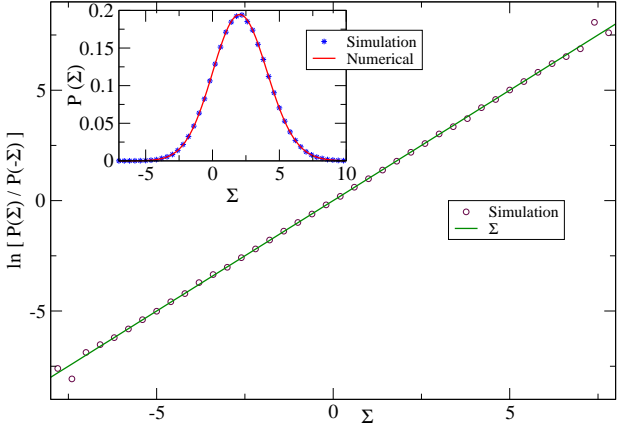}
\caption{(Color Online) Shows that $P(\Sigma)$ for molecular refrigerator model verifies the symmetry in eq.~(\ref{DFT5}). Inset compares $P(\Sigma)$ obtained from direct 
simulation with the same obtained by numerical inverse Laplace transform of $Z(u)$ in eq.~(\ref{z(u)}).Parameters for this plot are : 
($\gamma=\gamma'=1.0,~T=0.6$ and $\alpha_0=0.8$)}
\label{fig3}
\end{figure}
The joint probability of ${\bf{v}}$ and ${\bf{v'}}$ in the forward process is given by 
\begin{eqnarray}
&& \mathcal{P}_F({\bf{v}},{\bf{v'}}) = \mathcal{N} exp[-\frac{1}{4\gamma T } \sum_{i=0}^{N-1}\int_{t_i}^{t_{i+1}}(m\dot{v} + \gamma v + \gamma' v_i')^2 dt] \nonumber \\
&&\times~ exp[ -\frac{1}{2\Delta^2 \delta t}\sum_{i=0}^{N-1}(v_i'-v_i)^2\delta t] \times P_{eq}(v_0),
\end{eqnarray}
where $P_{eq}(v_0)= \sqrt{\frac{m}{2\pi T} }e^{-\frac{mv_0^2}{2T}}$ and $\mathcal{N}$ is normalization constant. 
If we take $\delta t\to 0$ and $\Delta^2 \to \infty$ such a way that $\delta t \Delta^2$ goes to a finite value, say $2\alpha_0$, then the exponent in the above equation can be expressed as a continuous stochastic integral which will 
correspond to the following Langevin equation
\begin{eqnarray}
m\dot{v}&=& - \gamma v - \gamma' v'  +\eta(t)~;~v'= v+ \frac{\xi(t)}{\gamma'}, \label{Blangeq1}
\end{eqnarray}
where, $\xi(t)$ is another mean zero white Gaussian noise with $\langle \xi(t) \xi(t') \rangle = 2\alpha_0 \gamma'^2 \delta(t-t')$. Because of the chosen Gaussian nature of the uncorrelated measurement process, 
the feedback control acts effectively 
as an extra white noise added to the original noise in the continuous limit. One can consider this additional noise $\xi(t)$ to be coming from another reservoir of temperature $T'=\gamma' \alpha_0$.

To check the validity of relations like eq.~(\ref{DFT5}) or (\ref{gDFT}), we need to consider the ratio of probabilities of a path in the forward process and its conjugate path in the reverse process. The reverse process is 
constructed as described in Sec.~(\ref{deriv}). Assuming the system starts from $P_{eq}(\bar{v}_0)$, we make measurements at each time $\bar{t}_i=\tau-t_i$ and get outcome $\bar{v}'_i$ with probability $p(\bar{v}'_i|\bar{v}_i)$, where 
$\bar{v}_i$ is the actual velocity at time $\bar{t}_i$ in the reverse process. To reach the continuous feedback limit we take the earlier limits in $\delta t$ and $\Delta^2$.  
We, now, take the ratio of the probabilities of the path $({\bf{v}},{\bf{v'}})$ in the forward process and it's conjugate path ($\bar{v}(t)=-v(\tau-t)~;~\bar{v}_i'=v_{N-i-1}'$) in the reverse process and compare this with eqs.~
(\ref{DFT4agn}), (\ref{entro-def}) and (\ref{feed-entro}) to get the different contributions of entropy production in $\Sigma$. Finally, adding all these contributions we get 
%\begin{eqnarray}
$\frac{\mathcal{P}_F({\bf{v}},{\bf{v'}})}{\mathcal{P}_R(\bar{\bf{v}},\bar{\bf{v}}')}= e^{\Sigma}, ~\text{with}~
 \Sigma = -c\int_0^{\tau} (\gamma'v^2 + \xi v) dt,$
%\end{eqnarray}
where $c=(\frac{1}{T}-\frac{1}{T'})$. 
Since $\Sigma$ is a non-Gaussian variable, it is very difficult to find its distribution in general. Instead we would first like to calculate characteristic function $Z(u)=\langle e^{-u\Sigma} \rangle$ in the large $\tau$ limit and 
then we will take inverse Laplace transform of $Z(u)$ numerically to get $P(\Sigma)$. It is more useful to first consider the restricted characteristic function $Z(u,\tau,v|v_0)=\langle e^{-u\Sigma} \rangle_{v,v_0}$, where the expectation is taken over all trajectories of the system which evolve from a given initial velocity $v_0$ to
a final velocity $v$ in time $\tau$. To evaluate $Z(u,\tau,v|v_0)$, we follow the procedure described in \cite{anu11} and we get,
\begin{eqnarray}
 Z(u,\tau,v|v_0) &=& \frac{e^{\tau\mu(u)}}{\sqrt{2\pi L_0(u)}}~e^{-\frac{1}{2}L_1(u)v^2}~e^{-\frac{1}{2}L_2(u)v_0^2},~ \nonumber \\
\text{where},~\mu(u) &=& \frac{\tilde{\gamma}}{2m}\bigl[ 1-\sqrt{1+\frac{a(u)^2}{\tilde{\gamma}^2}}\bigr] \label{Zmu}
\end{eqnarray}
 with  
$\tilde{\gamma}=\gamma + \gamma'$ and $a(u)=\sqrt{4 c^2 \gamma \gamma' T T' u(1-u)}$. The $L$ functions, present in the above equation, are given as \cite{anu11} :
\begin{eqnarray}
L_0(u)&=&\frac{\gamma T + \gamma' T'}{m\sqrt{\tilde{\gamma}^2+a^2}}, \\
L_{1,2}(u)&=& \pm \frac{m \gamma' T'}{\gamma T + \gamma' T'}\bigl[c u + \frac{1}{2 \gamma' T'}(\tilde{\gamma} \pm \sqrt{\tilde{\gamma}^2+a^2})\bigr]. \nonumber
\end{eqnarray}
%\nonumber

The characteristic function $Z(u)$ is obtained from $Z(u,\tau,v|v_0)$ through : $Z(u)=\int dv_0 \int dv ~Z(u,\tau,v|v_0)~P_{eq}(v_0)$. After performing the integration and some algebraic manipulation we have 
the following form of $Z(u)$ :
\begin{eqnarray}
Z(u)=  \frac{2e^{\tau \mu(u)}}{\sqrt{T}}~\sqrt{\frac{\sqrt{\tilde{\gamma}^2+a(u)^2}}{{\frac{a(u)^2}{\gamma T} + 2\frac{(\tilde{\gamma}+\sqrt{\tilde{\gamma}^2+a(u)^2})}{T}}}}. \label{z(u)}
\end{eqnarray}
From the expression of $a(u)$, it is clear that $\mu(u)=\mu(1-u)$ and $Z(u)=Z(1-u)$. 
Distribution $P(\Sigma)$ of $\Sigma$ can be obtained by taking inverse Laplace transform of the above $Z(u)$ numerically. We also obtain $P(\Sigma)$ by directly simulating the Langevin equation in (\ref{Blangeq1}) using velocity Verlet algorithm. We compare the distributions obtained through numerical 
inverse Laplace transform and direct simulation in the inset of fig.~(\ref{fig3}) where, we see an excellent agreement. In the same figure we plot $ln(~\frac{P(\Sigma)}{P(-\Sigma)})$ vs. $\Sigma$ to see that it verifies the theorem in eq.~(\ref{DFT5}). In our knowledge this is a first attempt to study continuous feedback process with measurement errors in the context of fluctuation theorems and we think this will be useful in understanding the action of many 
biological systems which are continuously controlled via feedback mechanisms. 

\section{Conclusion}\label{conclu}
In the conclusion, we have considered a class of feedback controlled stochastic dynamics where one can construct a reverse process with closed loop feedback process \emph{i.e.} one can allow performing measurements in the reverse 
process without violating the causality in contrast to \cite{hor10}. As a result we get an extra entropy production due to the measurement process. When we take into account this entropy production along with system and 
medium entropy productions we get 
the total entropy of the "full" system satisfying the fluctuation theorems and consequently the second law of thermodynamics. Next step would be to extend this idea for a more general stochastic dynamics and get more insight about 
the entropy production $\Delta s_p$. This kind of feedback mechanism would be easier to treat analytically in the context of finding efficiency of nano-machines or of biological machines. 

For demonstrating this idea we studied two examples with two different type of feedback mechanisms in detail and verified fluctuation theorems in both cases. The kind of feedback mechanism discussed 
in this paper, can also be realized in experiments, like, studying single paramagnetic spin in feedback controlled magnetic field, or pulled Brownian particle. 

I would like to thank G. Verley, D. Lacoste and K. Mallick for useful discussions.

\appendix
\renewcommand{\thesection}{}
\section{}
Expressions of the functions $A(u)$, $B(u)$ and $C(u)$ present in eq.~(\ref{zofa}) are the following :
\begin{eqnarray}
A(u)&=&\frac{1}{\Delta^2+u\sigma_{\tau}^2} \bigl ( \frac{\Delta^2\sigma_{\tau}^2f_2^2u^2}{2}-uf_2\Delta^2g_0(\tau) \nonumber \\ 
&& -\frac{u g_0(\tau)^2}{2}-\frac{f_1^2(\Delta^2+u\sigma_{\tau}^2)}{2\sigma_y^2} \bigr ) \nonumber \\
C(u)&=& \frac{f_1}{\sigma_y^2} - \frac{ue^{-\frac{k\tau}{\gamma}}g_0(\tau)}{\Delta^2+u\sigma_{\tau}^2}  -\frac{uf_2\Delta^2e^{-\frac{k\tau}{\gamma}}}{\Delta^2+u\sigma_{\tau}^2} + uf_2 \nonumber \\
B(u)&=&\left (\frac{1}{2\sigma_y^2} - \frac{u}{2\Delta^2} + \frac{ue^{-\frac{2k\tau}{\gamma}}}{2(\Delta^2+u\sigma_{\tau}^2)} \right )
\end{eqnarray}
with $g_0(\tau) = \alpha(1-e^{-\frac{k\tau}{\gamma}})$ and $f_2=\frac{\Delta^2\alpha k-T}{T\Delta^2}$.

\end{document}